\documentclass{aa}  

\usepackage{graphicx}
\usepackage{txfonts}
\begin{document}

   \title{The GALEX Ultraviolet Virgo Cluster Survey (GUViCS). VII.:}

   \subtitle{BCG UV upturn and the FUV-NUV color up to redshift 0.35\thanks{Tables 2 to 5 are available in electronic form
at the CDS via anonymous ftp to cdsarc.u-strasbg.fr (130.79.128.5)
or via http://cdsweb.u-strasbg.fr/cgi-bin/qcat?J/A+A/}}

   \author{S. Boissier\inst{1}
          \and
          O. Cucciati\inst{2}
         \and
        A. Boselli\inst{1}
\and
S. Mei\inst{3,4,5}
\and
L. Ferrarese\inst{6}
          }

   \institute{Aix Marseille Univ, CNRS, LAM, Laboratoire d'Astrophysique de Marseille, Marseille, France     \\          
(\email{samuel.boissier@lam.fr})
        \and
INAF - Osservatorio Astronomico di Bologna,
via Gobetti 93/3,
40129 Bologna, Italy
\and
LERMA, Observatoire de Paris, PSL Research University, CNRS, Sorbonne Universit\'es, UPMC Univ. Paris 06, F-75014 Paris, France
\and
Universit\'{e} Paris Denis Diderot, Universit\'e Paris Sorbonne Cit\'e, 75205 Paris Cedex
13, France
\and
Jet Propulsion Laboratory, California Institute of Technology, Cahill Center for Astronomy \& Astrophysics,  Pasadena, California, USA
\and
National Research Council of Canada, Herzberg Astronomy and Astrophysics Program, 5071 West Saanich Road, Victoria, BC,
V9E 2E7, Canada
             }

   \date{2017}

 
 \abstract
   {At low redshift,  early-type galaxies often exhibit a rising flux with decreasing wavelength in the 1000-2500 \AA{} range, called ``UV upturn''. The origin of this phenomenon is debated, and its evolution with redshift is poorly constrained. The observed GALEX FUV-NUV color can be used to probe the UV upturn  approximately to  redshift   0.5.}
   {We  provide constraints on the existence of the UV upturn up to redshift $\sim$ 0.4 in the brightest cluster galaxies (BCG)  located behind the Virgo cluster, using data from the GUViCS survey.}
   {We estimate the GALEX far-UV (FUV) and near-UV (NUV) observed magnitudes for BCGs from the maxBCG catalog in the GUViCS fields. We increase the number of nonlocal  galaxies identified as BCGs
with GALEX photometry from a few tens of galaxies to 166 (64 when restricting this sample to relatively small error bars).
We also estimate a central color within a 20 arcsec aperture. By using the $r$-band luminosity from the maxBCG catalog, we can separate blue FUV-NUV due to recent star formation and candidate upturn cases. We use Lick indices to verify their similarity to redshift 0 upturn cases.}
   {We clearly detect a population of blue FUV-NUV BCGs in the redshift range 
0.10-0.35, vastly improving the existing constraints at these epochs by increasing the number of galaxies studied, and by exploring a redshift range with no previous data (beyond 0.2), spanning one more Gyr in the past. These galaxies bring new constraints that can help distinguish between assumptions concerning the stellar populations causing the UV upturn phenomenon.
The existence of a large number of UV upturns around redshift 0.25 favors the existence of a binary channel among the sources proposed in the literature.
}
   {}


   \maketitle
%

\section{Introduction}


\citet{code69} presented for the first time evidence of an excess of far-ultraviolet (FUV) light in the bulge of M31. 
The International Ultraviolet Explorer (IUE) 
observations of Ellipticals allowed astronomers to characterize  
this as ``UV upturn'', 
i.e., a rising flux with decreasing wavelengths from about 2500 \AA{} to 1000 \AA{}  \citep[e.g.,][]{bertola82}.
The UV upturn was found in the nearby universe in quiescent gas depleted ellipticals and has been associated to old stars \citep{oconnell99,ferguson99}.
This feature has also been found in other old stellar systems such as M32 \citep{brown04} or open clusters \citep{buson06, buzzoni12}. 
Empirical work  to find the actual source of the upturn included the analysis of color-magnitude diagrams \citep{brown98}, the detection of individual horizontal branch stars \citep{brown00}, or surface brightness fluctuations \citep{buzzoni08}.

Since the UV upturn is found in early-type galaxies characterized by old stellar populations, an effect of age can be expected. Observed correlations also suggest a role of metallicity \citep{faber83,burstein88}. However, these results have been extensively discussed (see the conflicting results in \citealt{deharveng02,rich05,boselli05, donas07}).

The recent work on absorption indices revealing old and young populations by \citet{lecras16} has showed that there is still a strong interest to understand the nature of UV upturn sources and their contribution to stellar populations as a whole.
From the point of view of the evolution of galaxies and the role of the environment, it is important to understand the UV emission associated with old stellar populations in early-type galaxies and to determine whether it is related  to the environment \citep[see][]{boselli14}.


\citet{hills71} suggested that the UV emission in M31 could be related to the presence of very hot stars. \citet{renzini86} discussed the possible candidates in the context of stellar population evolution. This included young stars, hot horizontal branch stars, post-AGB stars, and binaries.
Several theoretical works studied the UV emission of various types of stars during their advanced evolution phases \citep[e.g.,][]{greggio90,dorman93,dorman95,dcruz96}.
These works showed that their UV output is very sensitive to small differences in the assumptions made. \citet{greggio90} suggested that stellar evolution theory alone could not provide  the explanation of the UV upturn.

It is still generally believed that the UV upturn is related to extreme horizontal branch stars \citep[][and references therein]{brown04}.
These stars could be low mass helium burning stars having lost their hydrogen-rich envelope \citep[e.g.,][and references within]{han07}.
However, the precise stellar evolution producing hot low mass stars is still debated.
Recent models include single-star evolution with both  metal-poor and metal-rich populations \citep[e.g.,][]{park97,yi98} or models including the effect of binarity, with  stars losing their hydrogen envelopes during binary interactions 
\citep{han07}.


The various models proposed for the UV upturn sources predict drastic differences concerning the evolution of galaxy colors with redshift.
\citet{renzini86} and \citet{greggio90} already suggested that observations at a look-back time of a few Gyr should allow us to distinguish between possible sources. 
In the case of single-star origin, the UV upturn is expected to occur late because it is produced by evolved stars. If it is related to binaries, its apparition can be more progressive. 
Higher redshift observations are needed to distinguish these different evolutionary scenarios.

From the observational point of view, 
\citet{rich05} found little evolution up to redshift 0.2 in a sample of 172 red quiescent galaxies
(not restricted to the most massive or to central  clusters) obtained by cross-matching SDSS and GALEX results. \citet{lee05} and \citet{ree07} compared the observed FUV-V color of nearby ellipticals to the brightest ellipticals in 12 remote clusters up to 
redshift 0.2, and also compared  this color to a few previous works \citep[see][]{brown04} up to redshift 0.6. 
These results suggest that the FUV-V restframe color is bluer by about 0.8 mag around redshift 0.2 with respect to redshift 0, with a large dispersion at all redshifts. The color of the few galaxies at redshifts higher  than 0.2 is close to the average color at lower redshifts. \citet{donahue10} presented the evolution of the FUV-R color with redshift in 11 brightest cluster galaxies (BCGs) in the redshift range 0.05-0.2, detected in FUV, with little evolution. Direct studies of the evolution of the UV upturn with redshift are still limited to small samples and include only a few tens of objects when limited to BCGs.
Table 1 list the samples of early-type galaxies with FUV and near-ultraviolet (NUV) magnitudes from GALEX.

\begin{figure}
\includegraphics[width=8.5cm]{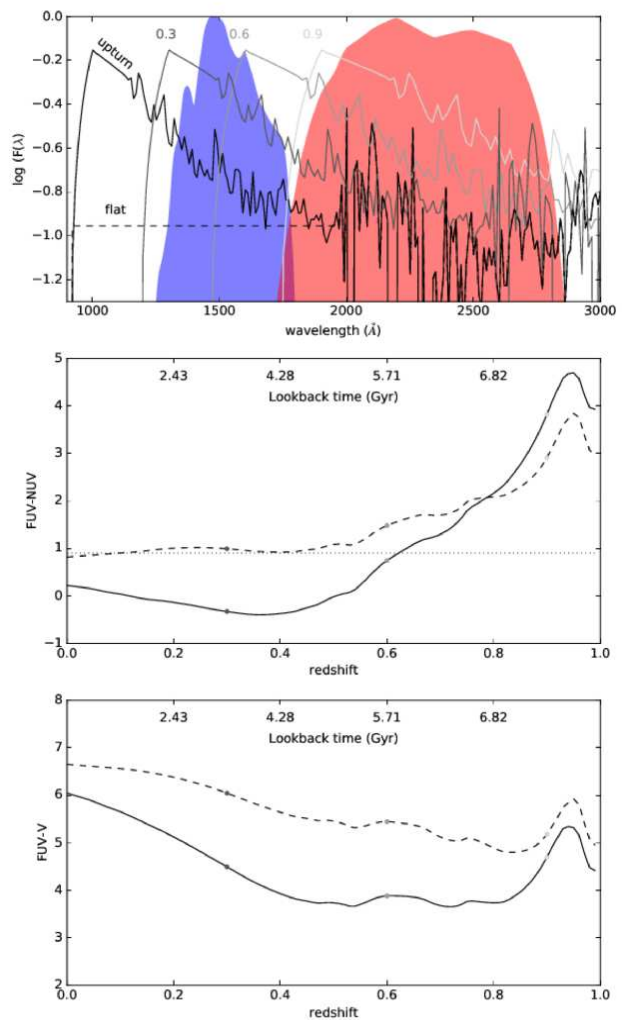}
\caption{Top: Arbitrarily scaled spectrum of NGC1399 (solid line). 
The spectrum was taken from the database of UV-Optical Spectra of Nearby Quiescent and Active Galaxies (http://www.stsci.edu/science/sed/sed.htm), 
the original UV data are from \citet{burstein88}. The spectrum was
smoothed for visualization, and extrapolated at low wavelength (between 912 and 1170 \AA) to match a typical 
upturn galaxy spectra \citep{yi98}. This spectrum is also shown after redshifting to z=0.3, 0.6, 0.9. 
A ``flat'' version is also shown (i.e., removing the upturn) for reference purposes (see text in the introduction).
The GALEX FUV and NUV passbands are respectively indicated as a blue and red shaded area. Middle: Evolution with redshift of the FUV-NUV color for the upturn and flat spectra as the solid and dashed curve, respectively. The dotted horizontal lines corresponds to FUV-NUV=0.9, the limiting color for an upturn as defined by \citet{yi11}.
Bottom: Evolution with redshift of the FUV-V color of the upturn and flat spectra as the solid and dashed curve, respectively.
}
\label{figupturn}
\label{figintro}
\end{figure}

\begin{table*}
\caption{UV upturn in early-type galaxy FUV-NUV samples}             
\label{tablesamples}      
\centering                          
\begin{tabular}{p{5cm} p{2.cm} p{2.5cm} p{2.cm} }      
\hline\hline                 
Reference & redshift range &  Galaxies  & Statistics \\    
\hline    
Rich et al. (2005)                                  & 0-0.2                 & early types  & 172 \\
Boselli et al. (2005)                               & 0                     & Virgo early types & 264 \\
Boselli et al. (2014)                               & 0                     & Virgo centrals    & 7 \\
Ree et al. (2007)                                   & 0-0.2                 & BCGs    & 12 \\
Donahue et al. (2010)                                & 0.06-0.18             & BCGs    & 10\tablefootmark{a} \\
Loubser et al. (2011)                               & 0                     & BCGs    &  36  \\
This work (all, FUV measurements) & 0.05-0.35             & BCGs     & 166 \\
This work (best sample\tablefootmark{b})                             & 0.05-0.35             & BCGs     & 64 \\
This work (confirmed upturn\tablefootmark{c})                        & 0.05-0.35             & BCGs     &  27 \\
\hline  
\end{tabular}
\tablefoot{
\tablefoottext{a}{Number of FUV detections.}
\tablefoottext{b}{Sample with low uncertainty on the central color (sum of uncertainties on each side lower than 1.2 mag) and excluding galaxies flagged as contaminated.}
\tablefoottext{c}{Red global NUV-$r$ color and blue central FUV-NUV color (see Sect. \ref{secyoungcompo}).}
}

\end{table*}


Several colors or other quantities have been used to detect and study the UV upturn. One of them is simply the FUV-NUV GALEX color that is easily accessible from GALEX data. It has been used  by, e.g.,  \citet{boselli05,donas07,loubser11}. This color probes the slope of the UV spectrum, and can be used to this end up to moderate redshift.
In Fig. \ref{figupturn} we show the spectrum of NGC1399, a Fornax elliptical with a strong UV upturn at redshift 0. We also show the spectrum shifted for a few redshifts, and the evolution of the corresponding FUV-NUV color as a function of redshift.
The FUV (around 1500 \AA) and NUV (around 2300 \AA) filters from GALEX are also indicated, showing that the FUV-NUV color probes the UV slope. The FUV-NUV color can thus be an indicator of the presence of an upturn, and of the value of the slope of the spectrum.
As a visual reference, we also show the evolution for a flat UV spectrum. Any upturn will be by definition bluer than this reference \citep[at redshift 0, it corresponds to a color close to the 0.9 limit, proposed by][to characterize the presence of an upturn]{yi11}. On the contrary, old stellar populations without upturn are redder than this reference.
The upturn FUV-NUV color is distinguishable from a flat spectrum  up to approximately redshift  0.5. Beyond this redshift, no more flux is found in the observed FUV band, and the color can no longer be used  to detect a UV upturn. 
The figure illustrates that the presence of a UV upturn can clearly be detected as a blue observed FUV-NUV color at the redshifts considered in this paper (below redshift 0.4).
In the nearby universe, the UV upturn is often studied on the basis of a color rather similar to FUV-V. Figure \ref{figupturn} shows  differences between the flat and upturn cases that are similar to those for the FUV-NUV color. For this color, however, we find a greater evolution of the observed FUV-V color (in AB magnitudes) with redshift. This makes it impossible to adopt a single color threshold for the detection of an upturn over the same redshift range. The two bottom panels of this figure can still help the reader  compare our study to this color choice. We note that this is very close to the classical (1550-V)$_{Burstein}$ from \citet{burstein88} with FUV-V (in the AB system) $\sim$ (1550-V)$_{Burstein}$ + 2.78 \citep{buzzoni12}.


The Virgo area was extensively studied in the FUV and NUV bands of GALEX in the context of the GUViCS project \citep{boselliguvics}. The photometry collected by GUViCS provides a deeper coverage than in most large areas over the sky. The UV properties of early-type galaxies inside the Virgo cluster were studied in \citet{boselli05}. 
In the present paper, we take advantage of these data to extract FUV and NUV photometry for massive galaxies in the background of the cluster, up to a redshift of about 0.35.

We select a sample of BCG galaxies from the maxBCG catalog  \citep{koester07}, extract FUV and NUV data for 177 of these galaxies from GUViCS, and perform a visual inspection to ensure the quality and noncontamination of these fluxes.
Considering the small statistics of existing BCG samples with FUV data  (e.g., only 36 galaxies in Loubser et al. 2011 at redshift 0; 12 in Rhee et al. 2007 up to redshift 0.2), 
even after removing the galaxies with possible contamination or large error bars, our sample brings new constraints for future models of the UV upturn population. We provide all our data in the form of easy-to-use tables for this purpose.

In Sect. \ref{secsample}, we present our sample and methods. 
The selection of galaxies showing upturn signs is discussed in Sect. 
\ref{secconfirmed}.
In Sect. \ref{secresults}, we show  the dependences of the FUV-NUV color  on luminosity and redshift in our sample, and discuss their implications.
A summary is given in Sect. \ref{secconclu}.

Throughout the paper, we use a flat cosmology (H$_0$=70, $\Omega_m$=0.3) to convert between look-back time $\tau$ and redshift ($z$). In our redshift range, the relation is linear with $\tau \sim 11.5 \times z$.

\section{Samples and data}
\label{secsample}

\subsection{ BCG sample}

In order to obtain a sample of galaxies that are as evolved as possible in the background of the Virgo cluster area, we extracted a sample of BCGs using the maxBCG catalog that
was computed from the Sloan Digital Sky Survey photometric data \citep{koester07}. 
We selected all galaxies with right ascension (RA)  in the 180-195 degrees range and 
with declination (DEC) in the 0-20 degrees range, and with 
available GALEX images in   the FUV and the NUV bands, from the GUViCS survey of this area \citep{boselliguvics}. 
This sample consists of 177 galaxies listed in Table \ref{tabsample}.
\citet{koester07} provided a number of properties of each BCG galaxy (position, redshift, luminosity) and of its cluster (e.g., number of members, luminosity of the members). 
We also performed a query to the DR13 SDSS release \citep{albareti16} to obtain the latest spectroscopic information. We checked that the spectroscopic redshifts are in agreement with the \citet{koester07} values for the 71 galaxies for which it was provided, and we increased the number of spectroscopic redshifts in this way up to 150 objects, i.e., the vast majority of our sample. We also obtained the SDSS spectroscopic class (``GALAXY'' for the 150 objects) and subclass based on line properties that result, in our sample, in one active galactic nucleus (AGN), three ``BROADLINE'' objects, one ``STAR-FORMING'' object.
Finally, SDSS also provides measurements of the Mg2 and H$\beta$ Lick indices often used
in the literature to study  the origin of the upturn in galaxies \citep[e.g.,][]{faber83,burstein88,boselli05,buzzoni12}. 


Table \ref{tabsample} compiles RA, DEC, photometric redshift, $r$- and $i$-band luminosity from the maxBCG catalog, 
while Table \ref{tabspec} provides the spectroscopic information obtained by querying the DR13 database: spectroscopic redshift, subclass, and Lick Mg2 and H$\beta$ indices when available.

\subsection{UV images}

The early-type  distant galaxies are often very faint in the ultraviolet bands, and a blind search can easily be affected by nearby objects (especially considering the $\sim$5 arcsec resolution of GALEX) or low signal-to-noise ratio. 
Due to the nature of the GUViCS survey and the
GALEX circular field of view, the survey is not homogeneous, and many
galaxies were observed on several occurrences.
We constructed stamps around the position of the BCGs by coadding any available UV images around our
sources. This was done using the Montage sofware \citep{jacob2010}
following the procedure described in \citet{boissier15}, which  allows the deepest possible UV exposure for each target. 
The UV original pixel is 1.5 arcsec wide, but since we reconstructed images from a variety of sources with arbitrary position shifts, we used Montage to project the UV images on a finer pixel grid. For practical reasons, we adopted the
same pixel scale (0.187 arcsec per pixel) as the optical data that we obtained from the NGVS survey 
\citep{ferrarese12}, as discussed in Sect. \ref{secoptical}.


\subsection{UV photometry}
\label{secuvphot}

Since we are targeting relatively small and faint galaxies whose shape 
in the UV is not known a priori, we computed photometry systematically 
in a number of circular rings around the BCG galaxies, with  
apertures of radii 20, 30, 50, 70, 90, 110, 130 pixels (with a size of 0.187 arcsecs), chosen to cover the range of sizes found in our 
sample. 
We note that the first aperture is only slightly larger than the GALEX PSF. We use it nevertheless because early-type galaxies are often very compact in the UV, and this allows us to have an estimate of the color even when a nearby galaxy could contaminate a larger aperture. The magnitude obtained is by definition partial. Aperture corrections for a point source at this size are 0.23 mag in both filters\footnote{http://www.galex.caltech.edu/researcher/techdoc-ch5.html}, thus the color is unchanged.
This central aperture has the advantage that it  often presents a higher S/N, and may be more sensible to UV upturn if the stellar populations giving rise to it are concentrated.
NGC1399,\label{secupturnconcentr} which we use to illustrate a typical upturn as seen in the nearby universe, clearly shows a FUV-NUV color gradient in the inner 30 arcsec, with the maximum upturn at the center as can be seen in Fig. 3 of \citet{gildepaz07}. 
The UV spectrum shown in Fig. 1 was obtained in the IUE aperture. For our distant objects, 
we certainly probe a larger physical size, thus this NGC1399 reference is likely to be the bluest color we can expect for a galaxy with a central upturn.

The sky value was measured in many independent regions around 
the galaxy and is void of obvious sources. The photometry 
and associated uncertainties were then computed as in 
\citet{gildepaz07}.
The photometry was corrected for the Galactic extinction, using the \citet{schlegel98} values for the visual 
extinction and $R_{FUV}$=8.24, $R_{NUV}$=8.20 \citep{wyder07}.

\subsection{Optical images}
\label{secoptical}

For all these galaxies, we fetched optical images from the NGVS survey 
\citep{ferrarese12}. Because the UV area we started from is larger than 
the NGVS area, we obtained these observations for about half of our sample (84 out of 177 galaxies). 
For the others, we fetched SDSS images. 
The optical 
images are not crucial for the analysis in this work, but were 
used for visual inspection
 allowing for instance to flag for possible contamination by nearby galaxies, or
signs of star formation in the form of spiral arms (Sect. \ref{secvisual}) or other 
morphological peculiarities.
In order to have homogeneous data; however, 
we use the SDSS photometry from the \citet{koester07} catalog for all our galaxies in the $i$ and $r$ band
(in order  to test the relations found in early-type galaxies and to test for the presence of 
a young stellar population). We do not perform photometry measurements in the optical images in this work.

\subsection{Visual inspection}
\label{secvisual}

A visual inspection of our images was performed. This step 
was important for this work for several reasons:

\begin{itemize}

\item[$\bullet$] We identified four BCGs with strong signs of star formation 
(spiral arms, prominent and spread out UV emission) that would obviously 
pollute any signal from the UV upturn; 

\item[$\bullet$] We identified objects for which the UV photometry was 
polluted by a nearby companion. 
In optical images, it is easy to spot small companions that   are unimportant in optical bands, but that can be the dominant source in the UV images if they are star-forming. Considering the GALEX PSF, in these cases the flux in the BCG region could be due to these companions and not to the BCG.
Comparing the images, we can easily say when the UV emission is centered on a companion rather than on the BCG;

\item[$\bullet$] We  chose the best circular aperture for this 
work. Having measurements in a collection of apertures, we could see in the image the surface covered by each aperture. We then selected the best one according to the following rules:
1) if possible the aperture including all the emission observed in the UV image  
(when  not possible, our magnitude was flagged as partial) and 
2) in any case, an aperture not polluted by a nearby companion 
(when  not the case, we used another flag to indicate contamination).
\end{itemize}

The inspection was performed independently by two people (O.C. and S.B.). 
A small discrepancy occurred for 25 \% of the galaxies (in most of the cases a  different optimal aperture was chosen with a different flag). The discrepancies were resolved through  discussion (usually adopting the smaller aperture to avoid possible pollution by a companion).

Table \ref{tabphoto} provides our results concerning the photometry, including the exposure time, the chosen aperture, the FUV and NUV magnitudes and their 1 $\sigma$ uncertainties. We have 11 galaxies without FUV measurements (measured flux below the sky level). In this case the table indicates a -99.9 magnitude and the -1 $\sigma$ column is replaced by the limiting magnitude as deduced from the sky noise measurement.

Table \ref{tabflags} provides the flags. For both FUV and NUV, a flag can be ``ok'' (the aperture encompasses all the observed emission in the image and there is no contamination), ``part.'' (we had to use a smaller aperture than the full observed emission to avoid contamination), or ``contam.'' (the flux is likely to be contaminated by a nearby source, usually a star-forming galaxy).
When possible, we preferred to have a part. flag, with a meaningful color in a small aperture, but in some cases, it was impossible to avoid a contamination.
We also added some notes that indicates clear signs of star formation and spiral arms (``spiral structure''), presence of arcs (``arc''), or presence of other signs that might be related to a merger or interactions (``shells/tails/asymetric/mergers''). These flags allow the identification of objects that may be affected for example by a recent merger or by star formation. It may be useful to distinguish them since, e.g., Using GALEX imaging, \citet{rampazzo11} found signs of star formation in their sample of 40 nearby early-type galaxies in low density  environments that can produce rings or arm-like structures. At high redshift (0.2 to 0.9),  \citet{donahue15} found in their CLASH BCG sample that BCGs with star formation activity indeed show  perturbed morphology, while quiet BCGs have a smooth aspect.

We found similar percentages of the various flags on the  subsamples with deep NGVS or SDSS optical images. Our flags are thus not affected by the source of the optical image that was examined together with our UV images. The only exception is the arc flag. We visually recognized four arcs, all of them in the NGVS images. Deep high quality exposures are necessary to recognize this feature. In our figures, we identify the galaxies flagged for these different categories so that it is possible to see if the one where they are found present differences in a systematic manner with respect to the global sample or not.

\begin{figure}
\includegraphics[width=9cm]{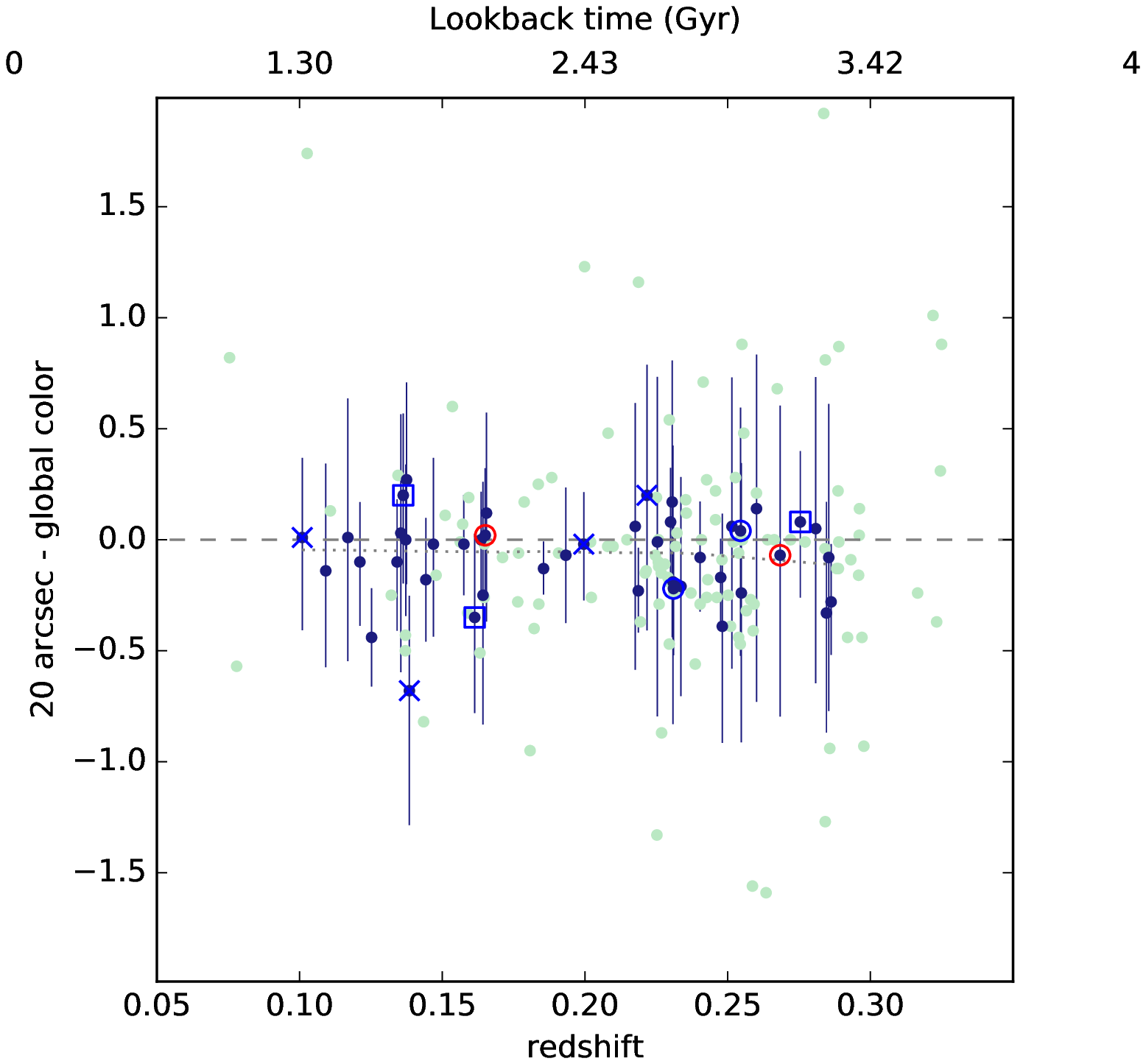}
\caption{Difference between the central and global FUV-NUV color.
Galaxies with contamination or very large error bars (1 $\sigma$ interval on the central or global color larger than 1.2 mag) are indicated by a pale green circle. 
The dark blue symbols with error bars are all our other points, including indications from the visual inspection results (see Table 4 and text for details). In blue: obvious spiral structure (crosses); 
arcs (circles); strong asymmetries, shells, tails, or signs of merger (squares). The red circles indicate galaxies with the SDSS spectral subclass BROADLINE.  The dashed lines indicates 0 (no color gradient) and the dotted line is a LOWESS fit  to the dark blue points.}
\label{figcolcol}
\end{figure}

\subsection{Aperture issues}

We also provide in Table \ref{tabphoto} the central color, i.e., the color of the innermost aperture. It provides an idea of the central upturn in the case of an extended object, where the population responsible for the upturn might be more concentrated (see Sect. \ref{secupturnconcentr}). 
When it was not possible to measure a FUV-NUV color in this aperture, the table indicates -99.9 color. 
We found that the central color correlates quite well with the total color. 
Figure \ref{figcolcol} shows the difference between the central and global color as a function of redshift. Over the range of redshift considered, for an intrinsic constant size, the observed size may change by a factor of about 2 with redshift. 
The difference in the selected optimal size compensates for this effect. For galaxies with ok flags, we selected mostly the 30- or 50-pixel apertures.
We chose  50 pixels  for most of the nearby galaxies ($z<$0.15), and  30 pixels  for all the distant ones ($z>$0.25).
Figure \ref{figcolcol} shows that we do not introduce a color trend by adopting the central color with respect to the global one. The central color has in general smaller error bars  than the global color. The uncertainty is on average reduced by a factor of 3 when using the 20 arcsec aperture with respect to the total aperture. 
For this reason, in the following we perform an analysis of the UV upturn in this smallest aperture. This has two main advantages.
First,  by definition, our color does not correspond to the total galaxy, but when we can have a total galaxy, they correlate. We are thus not limited to galaxies without contamination in the outer part (i.e., we can use galaxies having only a partial magnitude).
Second,  the error bar is smaller, which allows us to better distinguish trends. For the same adopted limit on the error bar size, we obtain larger statistics. Of course, this is not adequate for all analysis, thus  Table \ref{tabflags} also provides  the total magnitude  measured as described above.

This allows us to obtain what we call the ``best sample'', i.e., 64 BCGs with a central FUV-NUV color with global uncertainty (sum of the error bars on each side) lower than 1.2 magnitudes, and not contaminated.

\subsection{Sample of local galaxies in GUViCS}
\label{seclocal}
A previous work on UV upturn in elliptical galaxies inside the Virgo cluster, based on GALEX data, was performed by \citet{boselli05}. We refer the reader to this work for a detailed analysis of local galaxies. We consider here a similar comparison sample of local galaxies in the Virgo cluster, using the most recent set of data \citep{boselli14}.
For all galaxies, the UV data have been taken from the GUViCS catalog of UV sources published in \citet{voyer14}. 
The optical data in the SDSS photometric bands 
\citep{abazajian09} 
have been taken, in order of preference, from the SDSS imaging of the
\textit{Herschel} Reference Survey \citep{boselli10}, published 
in \citet{cortese12}, or from \citet{consolandi16}.
Given the extended nature of all these nearby sources, all magnitudes have been taken from  imaging photometry of extended sources and thus are total magnitudes.

Among this sample, we identify the seven galaxies being central to subgroups in Virgo \citep{boselli14} that are probably more similar to our BCGs than the other early-type galaxies in the local sample.


\begin{figure}
\includegraphics[width=9cm]{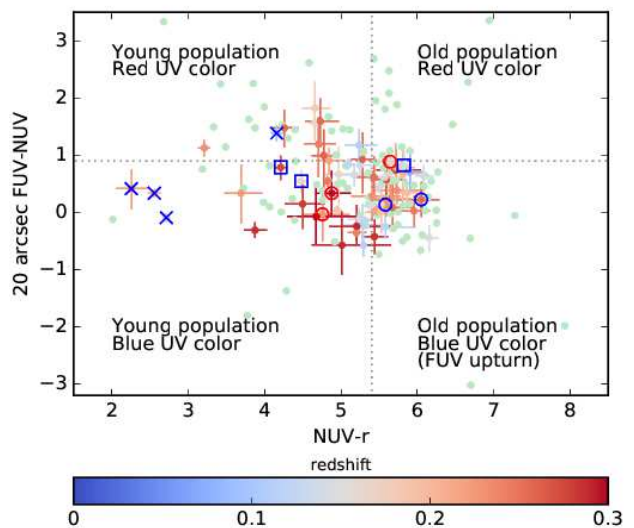}
\caption{Central FUV-NUV vs. global NUV-$r$ color-color diagram in the best sample (colored by redshift). For comparison, pale green dots show  the location of the other BCGs. The vertical line indicates NUV-$r$=5.4, above which there should be no contamination from young stars, and the horizontal line indicates FUV-NUV=0.9, below which the UV slope is consistent with a UV upturn \citep{yi11}.
Peculiarities found in our visual inspection (blue) or in the SDSS spectral subclass (red) are marked as in Fig. \ref{figcolcol}.}
\label{fignuvr}
\end{figure}

\section{Confirmed upturn sample}
\label{secconfirmed}

\subsection{Using optical photometry to exclude recent star formation}
\label{secyoungcompo}

Brightest cluster galaxies are not systematically quiescent systems: 10 \% to 30 \% of BCGs in optically selected samples show star formation or AGN activity \citep[][ and references therein]{donahue15}. A blue FUV-NUV is thus not necessarily the result of an evolved upturn population and can be the result of star formation activity. BCGs with star formation tend to have morphological signs such as filaments, elongated clumps, or knots \citep{donahue15}. From our visual inspection, we have identified a few obvious cases that can be excluded when focusing on the upturn phenomenon.
Fortunately,  \citet{yi11} and \citet{han07} showed that combining the FUV-NUV and NUV-$r$ (or FUV-$r$) colors allows the separation of upturns, compared to galaxies that are simply blue as a result  of young populations. 
\citet{donahue10} also showed that UV-optical colors are sensitive to even modest amounts of recent star formation.
We thus present in Fig. \ref{fignuvr} a FUV-NUV versus NUV-$r$ color diagram for our sample. The NUV-$r$ color is sensitive to young populations even if it does not probe exactly the same rest-frame waveband at all redshifts: blue colors indicate the presence of young stellar populations. 
We adopt the qualitative limits of \citet{yi11} to confirm the detection of rising UV flux when FUV-NUV is lower than 0.9; and the detection of a young population component when NUV-$r$ is bluer than 5.4 mag. The galaxies that were flagged for spiral structures all fall on the young population side of the NUV-$r$=5.4 limit, three of them being even among the bluest objects in  our galaxies.
The figure clearly shows that we have a number of BCGs in the best sample  at all redshifts  falling in the upturn/old stellar population part of this diagram. We consider that these 27 BCGs are likely to  present a UV upturn.

Another possible contribution to blue FUV-NUV color would be the presence of an AGN. \citet{oconnell99} however discusses the contribution of known bright nuclei in elliptical galaxies (M87, NGC4278). Only 10 \%
of the FUV luminosity is related to the nuclei. While it could affect the FUV-NUV color, this is marginal with respect to our observational uncertainties.
Moreover, the subclass from SDSS indicates one AGN and a few BROADLINE objects in 
our sample. Only three of them pass our   UV photometry criteria (uncertainties, noncontamination). These few objects do not distinguish themselves from the rest of our sample, as can be seen in the figures where we marked them. In summary, our results should not be affected by AGNs.

\begin{figure}
\includegraphics[width=9cm]{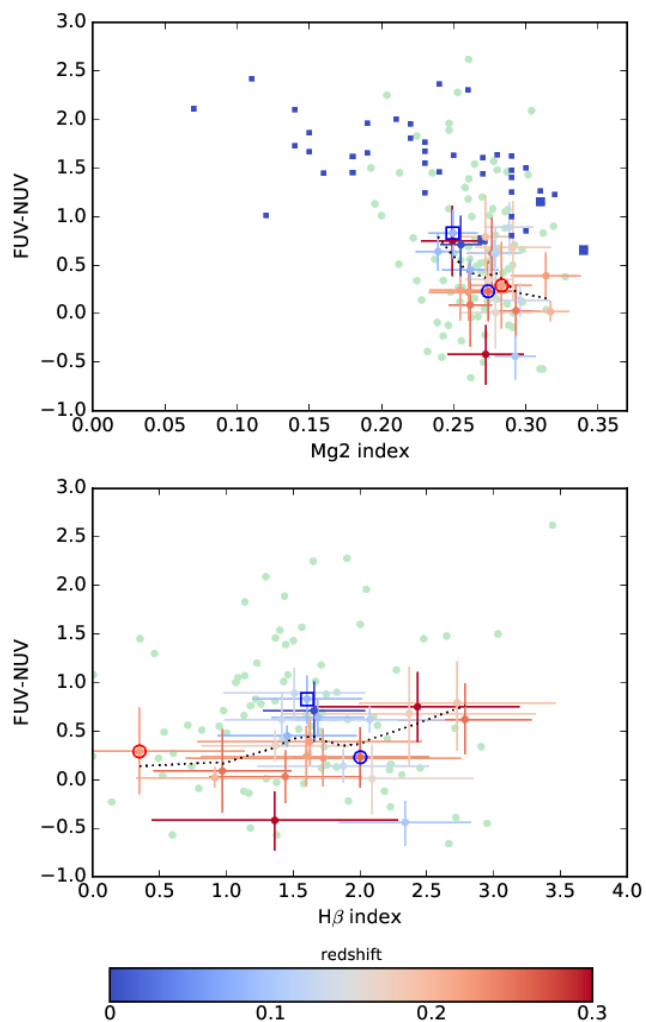}
\caption{FUV-NUV color as a function of the Mg2 and H$\beta$ Lick indices.
Galaxies not pertaining to our upturn sample are shown as pale green dots. For the confirmed upturn galaxies, the symbols are colored according to their redshift.
Peculiarities found in our visual inspection (blue) or in the SDSS spectral subclass (red) are marked as in Fig. \ref{figcolcol}. In the top panel, the squares show the relation found in the local sample (larger squares for centrals).}
\label{figindex}
\end{figure}

\begin{figure*}
\includegraphics[width=18cm]{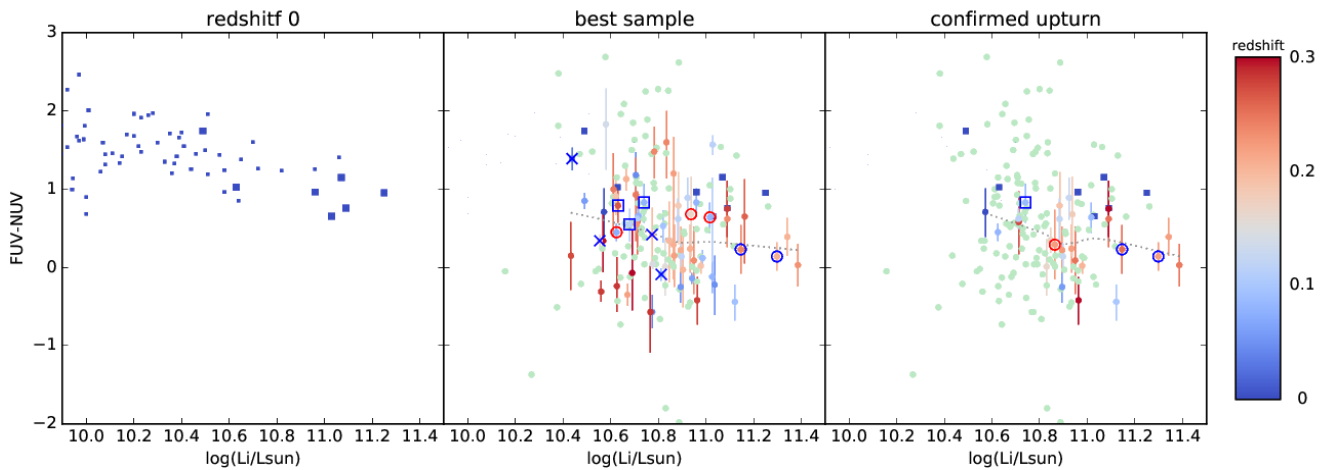}
\caption{FUV-NUV color as a function of the $i$-band luminosity. 
The left panel shows the relation for the redshift 0 early-type 
galaxies (Sect. \ref{seclocal}).
The central galaxies of the various subgroups in Virgo are indicated by larger symbols than the rest of the sample. These centrals are repeated in the other panels (squares).
In the middle panel, we show our full and best sample (colored circles). We use the central FUV-NUV color (smaller error bar, larger statistics). Galaxies with contamination or very large error bars (1 $\sigma$ interval larger than 1.2 mag) are indicated by a pale green circle. For the other galaxies, the color corresponds to the redshift, as indicated in the color bar. 
Peculiarities found in our visual inspection (blue) or in the SDSS spectral subclass (red) are marked as in Fig. \ref{figcolcol}. The dotted line is a LOWESS fit to our sample. In the right panel, the colored circles are used only for the galaxies with confirmed upturn.}
\label{figlum}
\end{figure*}

\subsection{Spectroscopic confirmation}

We show in Fig. \ref{figindex} the SDSS Mg2 and H$\beta$ Lick indices of
the ``confirmed upturn'' sample (and of the full sample for comparison).
These features have been used in the context of UV upturn studies since \citet{faber83} and \citet{burstein88}. SDSS provides values for many of our galaxies and the majority of our upturn sample (24 out of 27). 
In local galaxies, a trend between the Mg2 index and the strength of the upturn was  found, with stronger upturns in more metallic galaxies. With our sample of BCG galaxies, we probe only the most massive of the galaxies with respect to the local sample. We thus do not find a trend with the Mg2 index as in the redshift 0 sample from \citet{boselli05}. Our confirmed upturn sample behaves as the local  ``centrals''. As can be expected from color evolution of Fig. \ref{figintro}, the BCGs are slightly bluer at higher redshift, but present similar Mg2 values to the more massive of the local early-type galaxies showing upturns.

In the nearby universe, the H$\beta$ index has been used to distinguish galaxies with real upturn and those presenting residual star formation. The bottom panel of Fig. \ref{figindex} shows its value for our full sample and confirmed upturns. We find a very mild trend of bluer UV colors for smaller H$\beta$ indices. This pattern is typical of upturns \citep{buzzoni12}. Blue UV colors related to star formation are instead  found with higher values of the H$\beta$ index (above 2 $\AA$). Most of our BCGs are found with values between 1 and 2 \AA, typical of the passive galaxies with UV upturns.

From this section, we conclude that even if our selection of confirmed upturn were based on photometry alone, the spectroscopic information would confirm the status of these galaxies as quiescent with a real UV upturn, similar to that observed in the Local Universe and not polluted by residual star formation.

\section{Results}
\label{secresults}




\subsection{FUV-NUV color vs. luminosity}

We compare in Fig. \ref{figlum} the FUV-NUV color and $i$-band luminosity of the local sample, our BCG sample, and the upturn sample. While some works have found an evolution with luminosity \citep{boselli05}, the color in our BCG sample varies little with luminosity. However, the sample covers a small range of $i$-band luminosity, since BCG have by definition higher masses (well traced by the $i$-band luminosity) than the general population of galaxies.
The dispersion is reduced, however,  when we restrict the sample to galaxies with confirmed upturn with a very mild trend.

The central galaxies in subgroups of Virgo are found in the same range of luminosities as our BCGs. In this range, the upturn sample is slightly bluer on average than the local sample, in agreement with the K-correction that can be deduced from Fig. \ref{figintro}.
\citet{loubser11} studied a sample of 36 nearby BCGs. Their FUV-NUV color (0.79 $\pm$ 0.055) is bluer than normal ellipticals of the same mass, but they do not find a strong dependence on mass or other parameters within BCGs. This is similar to what we find at  higher redshift on average.

\begin{figure}
\includegraphics[width=9cm]{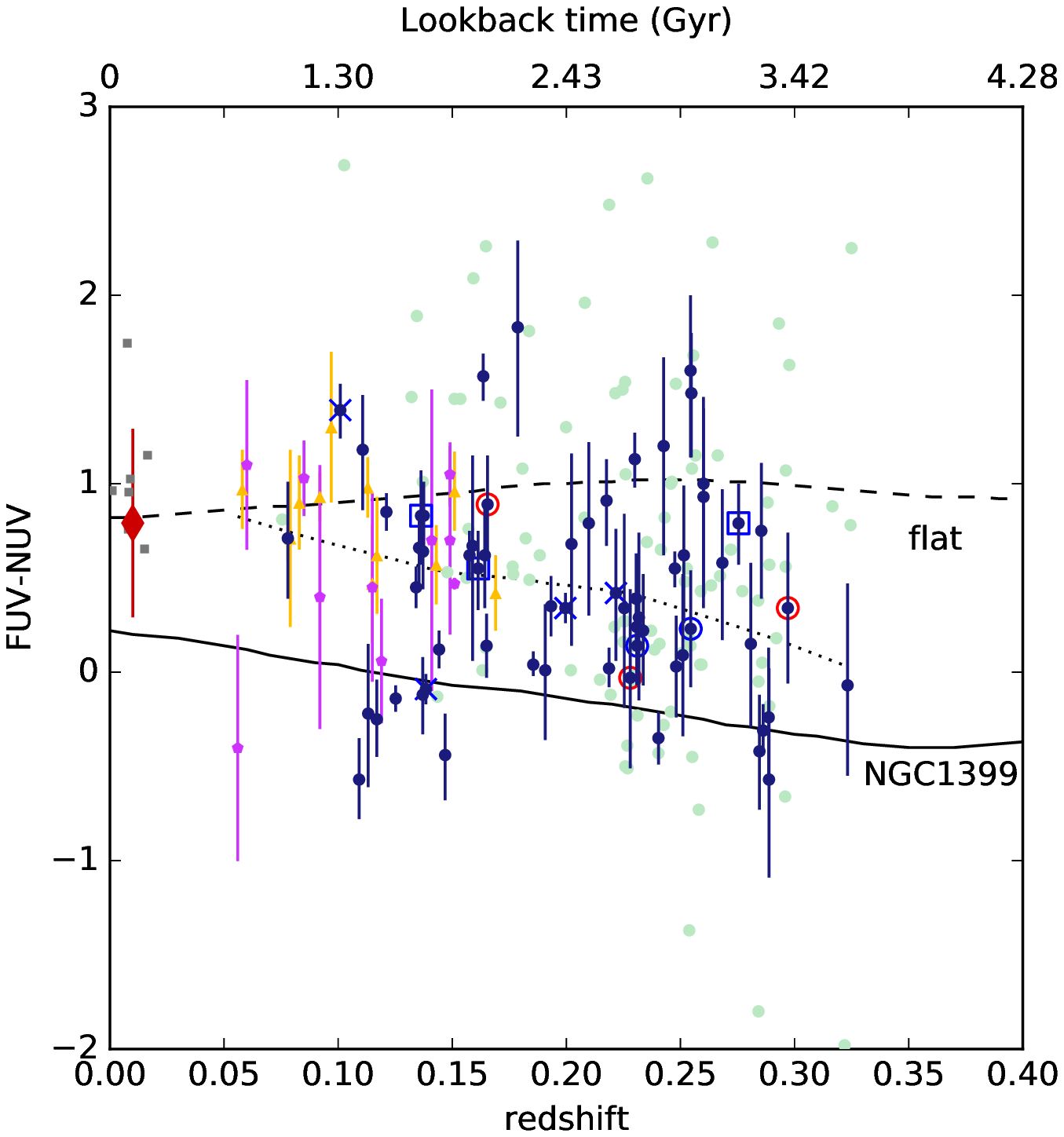}
\caption{Observed FUV-NUV color as a function of redshift. 
The gray squares are the redshift 0 centrals of Virgo subgroups  (Sect. 
\ref{seclocal}). The diamond is the average value of \citet{loubser11}, the error bar corresponding to the range of values they found for local BCGs. Intermediate redshift BCGs of \citet{ree07} and \citet{donahue10} are shown as orange triangles and magenta pentagons, respectively.
The circles correspond to our sample for which we use the central FUV-NUV color (smaller error bar, larger statistics). Galaxies with contamination or very large error bars (1 $\sigma$ interval larger than 1.2 mag) are indicated by  pale green circles, others by  dark blue circles with  corresponding error bars. 
Peculiarities are marked as in Fig. \ref{figcolcol} (noted in our visual inspection in blue; noted in the SDSS spectral subclass in red).
The solid (dashed) curve show the FUV-NUV color for the upturn spectrum of NGC1399 (for a flat spectrum) as a function of redshift.
A LOWESS fit to  our sample combined with the points of \citet{donahue10} and \citet{ree07} is indicated as the dotted line.
}
\label{figredshift}
\end{figure}

\subsection{FUV-NUV color vs. redshift}

As seen in the Introduction, the evolution with redshift of the UV upturn can bring direct constraints on the nature of stars producing it.
\citet{rich05} did not find any evolution of the UV upturn up to redshift 0.2. \citet{brown04} suggested that the UV upturn fades progressively with redshift up to redshift 0.6.  \citet{ree07} compared the observed FUV-V color of nearby ellipticals to the  ellipticals in 12 remote clusters up to redshift 0.2, and also compared  this color to six objects in two clusters at redshifts around 0.3 and 0.5, suggesting a weak evolution of this color (they did not show the evolution of the FUV-NUV color, but provided the corresponding data in their table).
\citet{lecras16} has suggested that the UV upturn appears at redshift 1 in massive galaxies and becomes more frequent at lower redshift. Their work is based on the fitting of line indices in synthesis population models where they can include a UV upturn component. They found that the rate of galaxies better fitted with models including an upturn is of 40 \% at redshift 0.6 and 25 \% at redshift 1.
A weak evolution could be consistent with the binary model of \citet{han07}, but
the constraints are still scarce, and our sample can bring new information.

Figure \ref{figredshift} shows our measurement of the observed FUV-NUV color as a function of redshift for the BCG sample. Here, we use the best  sample, i.e., all the galaxies with good constraints on the FUV-NUV color, which can be directly  compared to published works with similar data. In the next section, we  focus instead on the confirmed upturn that can be defined using optical photometry.
Our galaxies are compared to the other samples of BCGs with published FUV-NUV colors: the 36 local BCGs of 
\citet{loubser11} for which we indicate the average value and observed range of color, and at intermediate redshifts the BCG samples of \citet{ree07} and \citet{donahue10}. Our sample adds new points at redshift lower than 0.2, and brings unique measurements in the redshifts 0.2 to 0.3, namely a significant increment in look-back time. When a FUV-NUV color could be measured with a global range of uncertainty lower than 1.2 magnitude (this is an arbitrary value that we chose in order to balance precision and statistics), we do obtain relatively blue colors. 
Most of these galaxies are bluer than the flat spectrum that we used as an artificial reference, as expected in the case of an upturn.
A LOcally WEighted Scatterplot Smoothing (LOWESS) fit \citep[implemented
in Python by][]{cleveland79} 
performed on our BCGs combined with the two intermediate redshift samples stresses the trend of obtaining bluer colors at  higher redshifts. However, this blue color is not necessarily a sign of a UV upturn, as discussed above. We thus study the FUV-NUV color as a function of redshift for confirmed upturn in the next section.

\begin{figure}
\includegraphics[width=9cm]{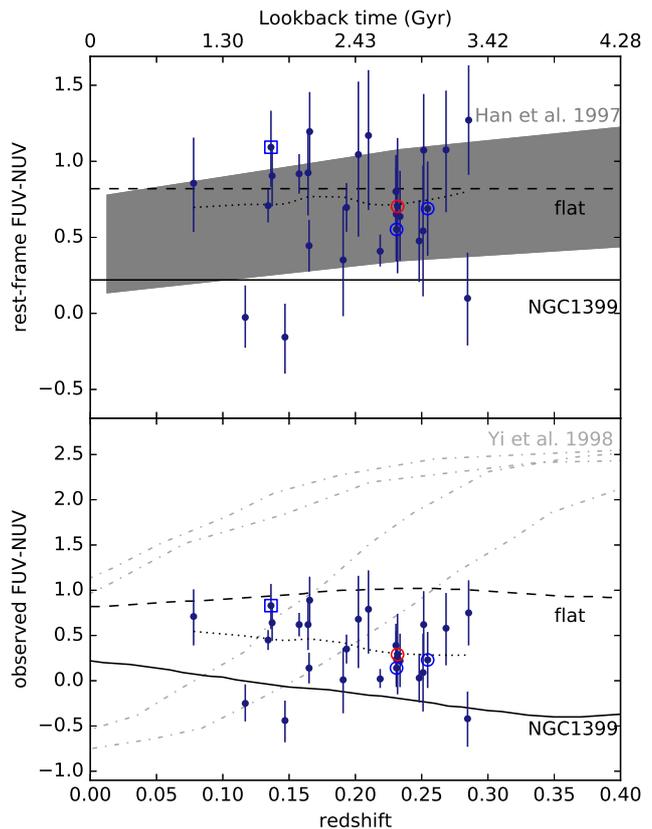}
\caption{FUV-NUV color as a function of redshift as in Fig. \ref{figredshift}, but keeping only our BCG galaxies with old populations and UV upturn, as defined by the bottom right part of Fig. \ref{fignuvr} (confirmed upturn sample). 
In the top panel, we show the rest-frame color. We
corrected the observed points assuming the NGC1399 spectrum since we selected upturn galaxies. In the bottom panel we show the observed color. In each panel our points are compared to model predictions.
The shaded area indicates the evolution of the rest-frame color for a SSP model of \citet{han07} including binaries (assuming a redshift 5 formation). The dot-dashed lines indicate the observed colors in typical models of \citet{yi98} based on single stars, for infall histories from \citet{tantalo96} for two galaxy masses (10$^{12}$ and 5 10$^{11}$ M$_{\odot}$), and two mass-loss efficiency parameter (0.7 for the two  lower and 1 for the two upper curves).
The dotted curve is a LOWESS fit to our data.
}
\label{figredshiftcut}
\end{figure}

\subsection{Detected UV upturn up to look-back times of 3.5 Gyr}
\label{secdiscu}


In Fig. \ref{figredshiftcut}, we finally show the FUV-NUV color as a function of redshift for our BCGs,  this time indicating only the BCGs for which a UV upturn is considered very likely based on the color--color diagram and the spectroscopic confirmation discussed in section \ref{secconfirmed}. 

We include in the figure the evolution predicted by different models.
At all our redshifts, we selected very massive passive galaxies. It is likely that their stellar mass or $i$ luminosity does not evolve much over the considered period since their star formation must have occurred at much earlier times in such galaxies \citep{thomas05}. The FUV-NUV evolution with redshift may thus be close to the actual color evolution of passively aging very massive galaxies. However,
we cannot be sure that we select the precursors of redshift 0 upturn galaxies when we select upturns at higher redshift. Thus, the evolution with redshift that we present is not necessarily the redshift occurring in any individual galaxy.
Even so, our results bring a new constraint for the stellar evolution models producing a UV upturn since models should at least predict the possibility of an upturn at the redshift when it is observed, which is not necessarily the case for all models.

The rest-frame FUV-NUV colors for the SSP models of \citet{han07} are shown in the top panel.
We computed rest-frame colors for our galaxies assuming the NGC1399 spectrum. 
In future model computations, it should be straightforward to obtain  the observed color to compare it directly to the values provided in our tables to avoid this step. 
The weak evolution they propose is quite consistent with the lack of evolution found in our sample
and the colors are globally consistent with these galaxies; 
we recall that  they do not represent the full population of BCGs, but those presenting a UV upturn, which is  a significant fraction (27 out of the 64 galaxies in the best sample).

We also show in the figure  four typical models among those presented in \citet{yi98} for two infall accretion histories and for two values of their mass-loss efficiency parameters. We refer to their paper for a more detailed description of their models. In this case, we reproduce their prediction for the observed color m(1500)-m(2500), which are not the GALEX bands but probe the UV slope in a similar wavelength range. These models tend to be characterized by more ample and rapid variations than the  \citet{han07} model is. Such large variation is not favored by our data (at least for a fraction of the BCG population).

We note that the models were basically constructed on the basis of redshift 0 constraints, with large uncertainties on the population responsible for the UV upturn. The inclusion of binary evolutions by \citet{han07} seems to help  reproduce the observed upturn at look-back  times of around 3 Gyr that we find in many of our galaxies.

\citet{ciocca17} recently found at redshift 1.4 blue color gradients in ellipticals in one cluster, with bluer UV-U colors in their center that may indicate the existence of a UV upturn population at even higher redshift.

\subsection{Environment influence on the UV upturn?}

We verified whether there is a correlation between the FUV-NUV color and other parameters in the maxBCG catalog. We could not find any significant trend, especially with those parameters related to the environment (e.g., number of cluster members). This is consistent with the findings of
\citet{yi11} and \citet{loubser11}, suggesting that the UV upturn is intrinsic to galaxies and not directly related to their environment.

\section{Conclusions}
\label{secconclu}

This work extends by a factor of several the size of BCG samples and provides constraints on their UV color. For the first time, we bring constraints on this subject to the redshift range 0.2-0.35, almost multiplying by 2 the look-back time with respect to previous studies. 

We took advantage of the GUViCS survey to study the FUV-NUV color (probing the UV slope) of 177 massive galaxies of the maxBCG catalog. Even if it is poorly constrained for many of them (due to their intrinsic faintness, and low exposure time in part of the GUViCS survey), we measured the FUV magnitude  for 166 objects in this sample.
Removing from this sample the galaxies with relatively large uncertainties and those with contamination, we obtained an interesting constraint (sufficient to distinguish different models) on the (central) FUV-NUV color of 64 BCGs at redshift 0.05 to 0.35.

Our most important result is that 27 out of the 64 BCGs with good photometry  at these redshifts present the characteristic of a UV upturn. They are selected on the basis of their blue FUV-NUV color, and of red optical colors suggesting an old underlying stellar population. The quiescent nature of these galaxies is confirmed by spectroscopic information.
%
They bring important constraints for models of the stellar populations responsible for the UV upturn phenomenon in very massive early-type galaxies. The comparison of our new data set with models from the literature favors a mild evolution with redshift like that obtained by models taking into account the effect of binaries on stellar evolution \citep{han07}. In conclusion, our data favors the existence of a binary channel to produce very hot stars that can produce a UV upturn, even up to redshift 0.35. 
This empirical work cannot give a definitive answer, however. Our tabulated data should offer a new constraint for future models of stellar evolution.

From the empirical point of view, follow-up work could be done to increase the statistics on the basis of extensive UV and optical data sets. Especially, a search for all massive galaxies, ellipticals, and BCGs in the NGVS optical catalog, and a similar systematic measurement of the UV color, but also of the UV-optical colors would be useful. 
In the long term, future large UV  facilities  (e.g., LUVOIR) could allow us to directly probe the UV spectrum of massive galaxies, providing more direct constraints on this still enigmatic phenomenon. 

\begin{acknowledgements}
This research is based on observations made with the NASA Galaxy Evolution Explorer. GALEX is operated for NASA by the California Institute of Technology under NASA contract NAS5-98034. We wish to thank the GALEX Time Allocation Committee for the generous allocation of time devoted to GUViCS. This research made use of Montage, funded by the National Aeronautics and Space Administration’s Earth Science Technology Office, Computation Technologies Project, under Cooperative Agreement Number NCC5-626 between NASA and the California Institute of Technology. Montage is maintained by the NASA/IPAC Infrared Science Archive. 
\end{acknowledgements}

%
%

   \bibliographystyle{aa} 
   \bibliography{ALLREFSedited.bib} 

\longtab{
\begin{longtable}{r r r r r r r r }
\caption{\label{tabsample}BCG galaxies in our sample.They are
selected from the reference catalog of \citet{koester07}, from which we list here
RA, DEC, photometric redshift, r and i band luminosities. }\\
\hline\hline
ID & RA (deg) & DEC (deg) & phot. redshift  & $L_r$ (10$^{10}$ $L_{\odot}$) & $L_i$ (10$^{10}$ $L_{\odot}$) \\
\hline
\endfirsthead
\caption{continued.}\\
\hline\hline
ID & RA (deg) & DEC (deg) & phot. redshift & $L_r$ (10$^{10}$ $L_{\odot}$) & $L_i$ (10$^{10}$ $L_{\odot}$) \\
\hline
\endhead
\hline
\endfoot
     BCG24 & 188.57278 &  9.766238 &  0.243050 &  17.90260 &  21.94700  \\ 
    BCG129 & 186.45179 &  7.971470 &  0.248450 &   9.48259 &  11.81850  \\ 
    BCG271 & 191.04632 & 14.482497 &  0.275450 &  11.28250 &  14.54510  \\ 
    BCG304 & 182.92986 &  8.018963 &  0.264650 &  18.48990 &  24.32040  \\ 
    BCG485 & 188.94962 & 15.555812 &  0.267350 &   7.44125 &   9.18799  \\ 
    BCG671 & 183.13768 &  6.076853 &  0.132350 &   8.40151 &  10.62250  \\ 
    BCG692 & 187.03522 &  8.945062 &  0.267350 &  16.49510 &  19.86410  \\ 
    BCG933 & 181.14637 & 11.095937 &  0.237650 &  11.01730 &  13.09370  \\
\end{longtable}

\begin{longtable}{r r r r r r r r }
\caption{\label{tabspec}Galaxies in our sample with spectroscopic information provided by SDSS:
redshift, sub-class (AGN, STARFORMING, or BROADLINE), Mg2 and H$\beta$ Lick indices and their uncertainty. }\\
\hline\hline
ID & spec. redshift  & sub-class & Mg2 & $\pm$ & H$\beta$ & $\pm$ \\
\hline
\endfirsthead
\caption{continued.}\\
\hline\hline
ID & spec. redshift  & sub-class & Mg2 & $\pm$ & H$\beta$ & $\pm$ \\
\hline
\endhead
\hline
\endfoot
    BCG24 &  0.230558 &            &  0.313916 &  0.024135 &  1.624143 &  0.842013 \\ 
    BCG129 &  0.252744 &            &  0.254789 &  0.018935 &  0.947647 &  0.611584 \\ 
    BCG271 &  0.272047 &            &  0.274995 &  0.017862 &  2.101470 &  0.584695 \\ 
    BCG304 &  0.248118 &            &  0.293060 &  0.016535 &  1.441411 &  0.571690 \\ 
    BCG485 &  0.284612 &            &  0.272066 &  0.026296 &  1.362648 &  0.921333 \\ 
    BCG671 &  0.137062 &            &  0.288143 &  0.015167 &  2.183753 &  0.400882 \\ 
    BCG692 &  0.231012 &            &           &           &           &           \\ 
    BCG933 &  0.225683 &            &  0.256044 &  0.019307 &  0.602522 &  0.693763 \\
\end{longtable}

\begin{longtable}{r r r r r r r r r r r r r}
\caption{\label{tabphoto}UV photometry measurements (see Sect. \ref{secuvphot}).}\\
\hline\hline
ID & FUV &  NUV          & Aperture & FUV      &  +1$\sigma$ & -1$\sigma$ & NUV &  +1$\sigma$ & -1$\sigma$  & FUV-NUV & +1$\sigma$ & -1$\sigma$ \\
   & exp.(s) &  exp.(s)  & (pixels)  &(AB) &   &  &(AB)&   &  & center &  &  \\
\hline
\endfirsthead
\caption{continued.}\\
\hline\hline
ID & FUV &  NUV          & Aperture & FUV      &  +1$\sigma$ & -1$\sigma$ & NUV &  +1$\sigma$ & -1$\sigma$  & FUV-NUV & +1$\sigma$ & -1$\sigma$ \\
   & exp.(s) &  exp.(s)  & (pixels)  &(AB) &   &  &(AB)&   &  & center & &  \\
\hline
\endhead
\hline
\endfoot
    BCG24 &    2315 &    3625 &     50 &  -17.50 &   0.23 &   0.29 &  -17.72 &   0.27 &   0.36 &   0.39 &   0.24 &   0.24 \\ 
    BCG129 &    1710 &    3060 &     30 &  -16.31 &   0.44 &   0.76 &  -16.58 &   0.35 &   0.52 &   0.55 &   0.71 &   0.87 \\ 
    BCG271 &    2568 &    5478 &     20 &  -16.20 &   0.32 &   0.46 &  -16.85 &   0.14 &   0.16 &   0.65 &   0.48 &   0.60 \\ 
    BCG304 &    1683 &    3101 &     30 &  -17.11 &   0.22 &   0.28 &  -17.53 &   0.17 &   0.21 &   0.03 &   0.27 &   0.27 \\ 
    BCG485 &    1746 &    4600 &     30 &  -17.15 &   0.22 &   0.28 &  -17.06 &   0.16 &   0.18 &  -0.42 &   0.30 &   0.31 \\ 
    BCG671 &     218 &    3457 &     50 &  -17.01 &   0.40 &   0.65 &  -17.39 &   0.15 &   0.18 &  -0.12 &   0.20 &   0.21 \\ 
    BCG692 &    4801 &    6313 &     30 &  -17.42 &   0.12 &   0.13 &  -17.78 &   0.11 &   0.12 &   0.14 &   0.18 &   0.18 \\ 
    BCG933 &    1295 &    1194 &     30 &  -16.41 &   0.28 &   0.37 &  -16.78 &   0.41 &   0.67 &   0.27 &   0.82 &   0.67 \\
\end{longtable}

\begin{longtable}{r l l l  }
\caption{\label{tabflags}Flags and notes for galaxies in our sample (see Sect. \ref{secvisual}).}\\
\hline\hline
ID & Flag FUV & Flag NUV & Note \\
\hline
\endfirsthead
\caption{continued.}\\
\hline\hline
ID & Flag FUV & Flag NUV & Note \\
\hline
\endhead
\hline
\endfoot
     BCG24 &         ok &         ok &                 \\ 
    BCG129 &         ok &         ok &                 \\ 
    BCG271 &      part. &      part. &                 \\ 
    BCG304 &         ok &         ok &                 \\ 
    BCG485 &         ok &         ok &                 \\ 
    BCG671 &         ok &         ok &                 \\ 
    BCG692 &         ok &         ok &            Arcs \\ 
    BCG933 &         ok &         ok &                 \\
\end{longtable}

}

\end{document}